\documentclass[psfig]{aa}  

\begin{document}

\title{Similarity solutions for radiation in time-dependent relativistic 
flows}


\author{L.B.Lucy}

\offprints{L.B.Lucy}

\institute{Astrophysics Group, Blackett Laboratory, Imperial College 
London, Prince Consort Road, London SW7 2AZ, UK}

\date{Received ; accepted }

\titlerunning{Similarity solutions}

\maketitle

\begin{abstract}

Exact analytic solutions are derived for radiation in time-dependent
relativistic flows. The flows are spherically-symmetric homologous explosions
or implosions of matter with a grey extinction coefficient. The solutions
are suitable for testing numerical transfer codes, and this is
illustrated for a fully relativistic Monte Carlo code.

\keywords{radiative transfer -- methods:analytical -- methods:numerical}

\end{abstract}

\section{Introduction}

In a recent paper (Lucy 2004), a Monte Carlo (MC) treatment accurate to
$O(v/c)$ of the
time-dependent transport of radiation in 3-D SNe is described and tested.
A major concern in that paper is establishing the accuracy of the MC code.
To this end, the 3-D code was applied to a 1-D problem that could be solved
independently with conventional numerical methods. Specifically, the test
problem was
to compute the bolometric light curve of a spherical SN in which
the transfer of UVOIR radiation is treated with a grey extinction
coefficient. 
An independent approach to this problem is provided by
Castor's (1972) co-moving frame (cmf) moment equations for
spherically-symmetric flows. The resulting
pair of partial differential
equations (PDEs) were solved with the Henyey method. 

	The final outcome of this test was entirely satisfactory: the mean
difference
between the two light curves is $\la 0.01$mag. for elapsed times $t$ from
$10$ to $50$ days. Nevertheless, the
differences were initially significant,
and it was not clear which code was in error. This question was eventually
answered by testing each code separately against an exact similarity solution
of Castor's equations. The cause of the differences could then be
traced to the poor spatial resolution of the MC code.

	In the interest of concise presentation, the essential part played
by this similarity solution was not described in the earlier paper
(Lucy 2004). But
subsequently, this solution was found to generalize
to all orders of $v/c$. Accordingly, since time-dependent relativistic flows 
and the associated transfer problems are of interest for such phenomena as
gamma-ray bursts and micro-quasars, this paper derives this solution (and
variants thereof) and illustrates its use in testing a relativistic
transfer code.

\section{Basic equations}

Similarity solutions will be sought for the frequency-integrated radiation
field in a homologously expanding or contracting flow. The configuration is 
spherically-symmetric, and the matter has grey extinction 
($\chi$) and 
integrated emissivity
($\eta$) coefficients that are isotropic in the cmf.

\subsection{The transfer equation}

Mihalas (1980) has derived the general time-dependent cmf transfer
equation for spherically-symmetric flows with relativistic velocities.
From this, he derives the zeroth and first frequency-integrated    
moment equations, whose earlier derivation by Prokof'ev (1962) is
acknowledged. When terms of $O(v^{2}/c^{2})$ and higher are neglected,
Castor's (1972) moment equations are recovered.

	Following Mihalas (1980), the interaction coefficients ($\chi ,\eta$)
and the
radiation field ($I,J,H,K$)
are expressed in the cmf but spacetime coordinates $(r,t)$ and flow velocities
$(v)$ are measured in the rest frame (rf). Since radiation quantities
always refer to the cmf, primes or suffixes to indicate this frame are
omitted (cf. Mihalas 1980, Sect. III).

	In a homologous spherical flow, we have $v = r/t$. For $t > 0$,
the flow is an explosion ($ v > 0$) starting with infinite density at $t = 0$.
For $t < 0$, the flow is an implosion ($ v < 0$) leading to infinite density
at $t = 0$.

	When Eq. (2.12) of Mihalas (1980) is applied to the 
homologous flow of matter with a grey extinction coefficient,
the resulting frequency-integrated transfer equation is 
\begin{eqnarray}
 \frac{\gamma}{c} (1+\beta \mu) \frac{d I}{d t} +
 \frac{\mu}{\gamma c t} \frac{\partial I}{\partial \beta} +
 \frac{\gamma}{\beta c t} (1-\mu^{2}) \frac{\partial I}{\partial \mu}
 + \frac{4 \gamma}{c t} I
 \nonumber \\
 = \eta - \chi \, I
\end{eqnarray}
Here the dependent variable is the integrated specific intensity 
$I(\mu; \beta,t)$, the radial coordinate $r$ has been replaced by
$\beta = v/c$, the time derivative is Lagrangian, and
$\gamma = 1/\sqrt{1-\beta^{2}}$. 

\subsection{Moment equations}

Moment equations can be derived from Eq. (1) as usual by multiplying
by $\frac{1}{2} \mu^{m}$ with $m = 0,1,2,...$ and then integrating over
direction
cosine $\mu$. Alternatively, Eqs. (2.16) and (2.17) in Mihalas (1980) can
be simplified to the case of homologous flow, as above for his transfer
equation.

	The resulting equations are 
\begin{eqnarray}
 \frac{\gamma}{c} \frac{d J}{d t} +
 \frac{\gamma \beta}{c} \frac{d H}{d t} +
 \frac{1}{\gamma c t} \frac{\partial H}{\partial \beta} +
 \frac{2 \gamma}{\beta c t} (H+2 \beta J)
 \nonumber \\
 = \eta - \chi J
\end{eqnarray}
for the zeroth moment, and
\begin{eqnarray}
 \frac{\gamma}{c} \frac{d H}{d t} +
 \frac{\gamma \beta}{c} \frac{d K}{d t} +
 \frac{1}{\gamma c t} \frac{\partial K}{\partial \beta} +
 \frac{\gamma}{\beta c t} (3K - J)
 \nonumber \\
 = - \chi H
\end{eqnarray}
for the first moment. 
   
	Following Mihalas (1980), Eddington's flux variable $H$ is here
preferred to the standard flux $F = 4 H$. The integrated physical
flux in the cmf is therefore ${\cal F} = 4 \pi H$.

	Comparing Eqs. (2) and (3) with Eqs. (2.16) and (2.17) of
Mihalas (1980), we see
that specializing to homologous flow has resulted in a huge simplification
in the coefficients of the moments $J$, $H$ and $K$. Most notably, the
coefficient of $K$ in the zeroth moment equation reduces to zero,
so now only
$J$ and $H$ appear in this equation.

\subsection{Separation of variables}

In order to construct simple solutions of these equations for use in testing
computer codes, we first effect a separation of variables. This is 
achieved by writing the specific intensity 
\begin{equation}
 I(\mu; \beta,t) = I_{1}(\mu; \beta) \, (t_{1}/t)^{p}
\end{equation}
where $t_{1}$ is an arbitrary reference time, and the exponent
$p$ is unspecified. With this assumption, each term on the left-hand
side of Eq. (1) scales as $(t_{1}/t)^{p+1}$. Accordingly, we must
assume that  
\begin{equation}
 \eta(\beta,t) = \eta_{1}(\beta) \, (t_{1}/t)^{p+1}
\end{equation}
and
\begin{equation}
 \chi(\beta,t) = \chi_{1}(\beta) \, (t_{1}/t)
\end{equation}
Not surprisingly, this necessary scaling of the extinction coefficient per
unit volume implies that the ratio of $1/\chi$,  the local mean free path  
of a photon,
to the radius of the configuration is time-independent. 

	When Eqs. (4)-(6) are substituted in Eq. (1), the scaling factor
cancels by construction, and we thus obtain the transfer equation
\begin{eqnarray}
 \frac{\mu}{\gamma c t_{1}} \frac{\partial I_{1}}{\partial \beta} +
 \frac{\gamma}{\beta c t_{1}} (1-\mu^{2}) \frac{\partial I_{1}}{\partial \mu}
 + \frac{\gamma}{c t_{1}}[4- p (1+ \beta \mu)]  I_{1}
 \nonumber \\
 = \eta_{1} - \chi_{1} \, I_{1}
\end{eqnarray}
that determines the scale-free radiation field $I_{1}(\mu; \beta)$.

	The moments $J$, $H$ and $K$ evidently also scale as 
$(t_{1}/t)^{p}$. The equations satisfied by the corresponding
scale-free moments $J_{1}(\beta)$, $H_{1}(\beta)$ and $K_{1}(\beta)$ are
\begin{eqnarray}
 \frac{1}{\gamma c t_{1}} \frac{d H_{1}}{d \beta} +
 \frac{\gamma}{\beta c t_{1}} (2- p \, \beta^{2}) \, H_{1} +
 \frac{\gamma}{c t_{1}}(4- p) \, J_{1}
 \nonumber \\
 = \eta_{1} - \chi_{1} J_{1}
\end{eqnarray}
for the zeroth moment, and
\begin{eqnarray}
 \frac{1}{\gamma c t_{1}} \frac{d K_{1}}{d \beta} -
 \frac{\gamma \beta}{c t_{1}} p \, K_{1} +
 \frac{\gamma}{\beta c t_{1}} \, (3 K_{1} - J_{1}) +
 \frac{\gamma}{c t_{1}} \, (4 - p) \, H_{1}
 \nonumber \\
 = - \chi_{1} H_{1} \; \; \; \; \;
\end{eqnarray}
for the first moment.

	Eqs. (7)-(9) are the basic equations of this
investigation.

\section{Testing a Monte Carlo code}

If MC techniques are used directly to simulate the physics of radiation
transport, then the MC quanta are photons and convergence to the solution
of the Radiative Transfer Equation (RTE) requires that
${\cal N} \rightarrow \infty$, where ${\cal N}$ is the 
number of photons whose interaction histories are followed. In such a code,
in addition to crossings of boundaries, MC quanta are created spontaneously
within the computational domain $D$
by sampling the thermal emissivity and may subsequently
be destroyed within $D$ by absorption.
Because the RTE is directly simulated, testing such MC codes is not
fundamentally different from testing a conventional numerical solution of the
RTE.

	However, for tranfer problems involving interactions between radiation
and the internal energy states of matter, there are advantages in taking the
MC quanta to be indestructible and indivisible energy (${\cal E}$-)
packets (Lucy 2003, and references therein). In such a code,
in addition to crossings of boundaries, ${\cal E}$-packets are created
spontaneously within $D$
by sampling the {\em net} emissivity (i.e., emission minus absorption)
but then, though 
the nature of the contained energy may change, they
are not subsequently destroyed within $D$ by absorption.

\subsection{Moment solution}

Let the integrated net emissivity per unit volume at time $t_{1}$
be $4 \pi \tilde{\eta}_{1}$, where 
\begin{equation}
 \tilde{\eta}_{1}(\beta) = \eta_{1} - \chi_{1} J_{1}  
\end{equation}
This quantity creates radiant energy within the configuration by
a physical mechanism that need not be specified for test problems.

	If we replace the right-hand side of Eq. (8) by $\tilde{\eta}_{1}$
and eliminate $J_{1}$ from the left-hand side by setting $p = 4$,
the result is an ordinary differential equation (ODE) for $H_{1}(\beta)$,
\begin{equation}
 \frac{d H_{1}}{d \beta} +
 \frac{2 \gamma^{2}}{\beta} (1-2 \beta^{2}) \, H_{1} 
 = \gamma c t_{1} \tilde{\eta}_{1} 
\end{equation}
This equation can be solved analytically with the integrating factor
$\beta^{2}/\gamma^{2}$. The solution satisfying the boundary condition
$H_{1}(0) = 0$ is
\begin{equation}
 H_{1}(\beta) = c t_{1} \, \frac{\gamma^{2}}{\beta^{2}}
      \int_{0}^{\beta}  \tilde{\eta}_{1}(b) \; b^{2} \sqrt{1 - b^{2}} \; d b 
\end{equation}

	Note that this formula is exact. In transfer theory, analytic
formulae for moments are typically not exact because they
are derived with Eddington's closure approximation $K = J/3$.
Here this is not necessary: K and J
drop out of the zeroth moment equation because of the assumptions
of homologous flow and scaling exponent $p = 4$.   

	A second point to note is that $H_{1}(\beta)$ has been derived
without specifying the scale-free extinction coefficient.
Accordingly, Eq. (12) is valid for arbitrary $\chi_{1}(\beta)$.

	It is of interest to note that the scaling $p = 4$ arises
naturally in the limiting case of completely opaque matter  
within which there
is no net emissivity. In this case, the right-hand side of Eq. (2)
is zero, and the cmf flux $H = 0$ since
radiation is position-coupled to matter. The solution of Eq. (2) is then
such that $J \propto t^{-4}$, corresponding to 
adiabatic evolution of the radiation energy density present initially.
In contrast,
in the solution derived here, this same scaling is maintained because every
layer's losses due to flux divergence is exactly replaced by the
net emissivity in that layer. Moreover, the similarity solution represents
the state reached when initial conditions have been erased.  

\subsection{A particular case}

The simplest case for testing a MC code is
when $\tilde{\eta}_{1}(\beta)$ is independent of $\beta$.
This also simplifies the evaluation of the exact $H_{1}(\beta)$
since the integral in Eq. (12) is then analytic.
The result is 
\begin{equation}
 H_{1}(\beta) = \frac{1}{8} \; c t_{1} \; \tilde{\eta}_{1} \;
  \frac{\gamma^{2}}{\beta^{2}} \;
  [\, sin^{-1} \beta  - \frac{\beta}{\gamma} \,(1-2 \beta^{2}) \, ]
\end{equation}
Thus we obtain an exact closed-form expression for the cmf flux in a
particular time-dependent relativistic flow.

	The corresponding moment $J_{1}(\beta)$ cannot be obtained 
exactly. But an approximate formula can be derived from
Eqs. (9) and (13) with the help of Eddington's closure approximation
and surface boundary condition. The resulting formula for $J_{1}$
can then be substituted in Eq. (10) to obtain an approximate formula
for the conventional emissivity $\eta_{1}$. Details are omitted.

\subsection{Monte Carlo calculation}

In order to illustrate how similarity solutions can be used to test
codes, we now briefly report MC calculations for relativistic homologous
flows. The MC code is a spherically-symmetric and fully relativistic version  
of the 3-D code described recently (Lucy 2004). As in that code, the
MC quanta are indestructible ${\cal E}$-packets. The calculations start at
reference time $t_{1}$ with no ${\cal E}$-packets present. But as time
advances, ${\cal E}$-packets spontaneously appear in accordance with the
net emissivity $\tilde{\eta}_{1} (t_{1}/t)^{5}$ and then propagate through
the configuration interacting with matter in accordance with extinction
coefficient $\chi_{1}(t_{1}/t)$.      

	We choose to create equal numbers of ${\cal E}$-packets in equal
intervals of $log \, t$ and having
cmf energies $\epsilon(t)$ that are independent of $\beta$. The energy
$dE$ created in the cmf within the space-time element $dV \, dt$ is
$dE = 4 \pi \tilde{\eta} dV \, dt$. But since $dV \, dt$ is invariant
under the
Lorentz transformation, we may regard $dV \, dt$ as referring to the rf.    
Accordingly, for the particular case (Sect. 3.2) where $\tilde{\eta}$ does not
depend on $\beta$, the
total cmf energy created in the rf interval $dt$ is  
$4 \pi \tilde{\eta} V \, dt$, where $V (t)= V(t_{1}) (t/t_{1})^{3}$
is the rf volume at time t. Thus, if 
$d {\cal N}/d \ell n t$ gives the packet creation rate, and we set $p = 4$,
then the  packets' cmf energies are
\begin{equation}
 \epsilon(t) = 4 \pi \eta_{1} t_{1} V_{1} \, \left(\frac{t_{1}}{t} \right)
        \left(\frac{d {\cal N}}{d \ell n t} \right)^{-1}
\end{equation}

	With $\tilde{\eta}$ independent of $\beta$, $dE \propto dV$ at fixed
rf $t$. Accordingly, since we choose $\epsilon$ to be independent of 
$\beta$, equal numbers of packets are created in equal
rf volume elements. The appropriate MC sampling algorithm for the initial
rf radius of a packet created at $t$ is $r = R_{*} \sqrt[3] z$, where
$R_{*} = c \beta_{*} t$ and $z$ is a random number from $(0,1)$.
Here $\beta_{*}$ is the value of $v/c$ at the surface.

	The transport of these packets through the configuration is
as described in the earlier paper (Lucy 2004) except that now
the factor $\gamma$ in the Doppler formula is restored to make the code
fully relativistic. The cmf energies of the packets that escape from the
surface in each rf
time step $\Delta t$ are summed and then divided by $\Delta t / \gamma_{*}$ 
to obtain an estimate of the cmf luminosity $L(t)$

\begin{figure}
\vspace{8.2cm}
\includegraphics{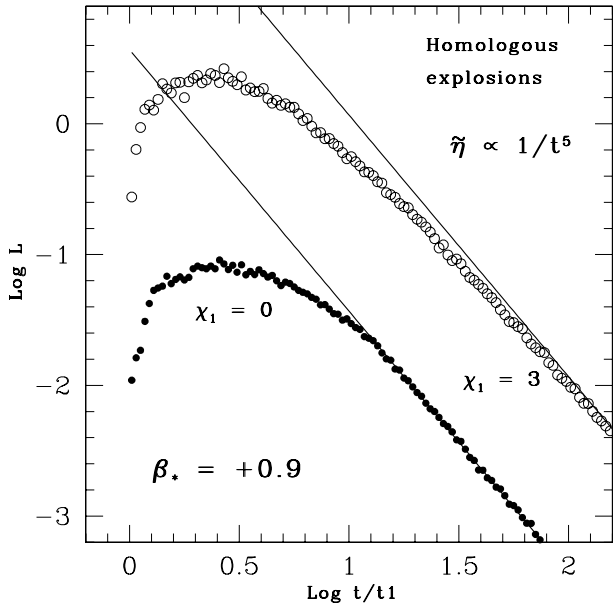}
\caption{Explosions. Comparison of MC calulations ({\em circles}) of the cmf
luminosity $L(t)$
with predictions of similarity theory ({\em straight lines}). The surface
velocity $v_{*} = 0.9 c$. The unit for the indicated scale-free 
extinction coefficients $\chi_{1}$ is $1/R_{1}$, where $R_{1} = v_{*} t_{1}$.
The unit of luminosity is $4/3 \pi R_{1}^{3} \times 4 \pi \tilde{\eta}_{1}$.
The luminosites for $\chi_{1} = 3$ have been increased by $1.5dex$.}
\end{figure}

	Monte Carlo simulations for two strongly relativistic explosions
are plotted in Fig. 1. For these simulations, the rf time steps are
$\Delta log \, t = 0.02$, during each of which ${\cal N} = 1000$ additional
${\cal E}$-packets are created with cmf energies given by Eq. (14).

The evolution of the
cmf luminosity is shown starting at $t_{1}$, together with the predicted
result
$L = 4 \pi R_{*}^{2} \times  4 \pi H_{1} (\beta_{*}) \times
(t_{1}/t)^{4}$, with $H_{1} (\beta_{*})$ from Eq. (13). One solution
is the free streaming limit ($\chi_{1} = 0$) and the second ($\chi_{1} = 3$)
has photon mean free paths $= R_{*}(t)/3$. In both cases, the MC 
solutions tend asymptotically to the similarity solution, the convergence
being somewhat slower for $\chi_{1} = 3$ because of the longer residence times
of the ${\cal E}$-packets. Since agreement is achieved in both cases, the
prediction, for $p = 4$, that the cmf flux $H$ is independent of $\chi_{1}$
is confirmed.

\subsection{Relativistic implosions}

Although relatively trivial, it is of interest to illustrate
the application to implosions. Accordingly, Fig. 2 shows
the same two test
problems but with the signs of $t_{1}$ and $\beta{*}$ reversed. Again,
after a transition period during which the internal radiation field is
established, the MC solutions converge to the similarity solution.

	This possibility of testing numerical treatments of radiative
transfer
in relativistic implosions is possibly relevant for codes that simulate
laser-driven implosions in support of the quest for fusion by 
inertial confinement.

	Accurate numerical solutions for relativistic inflows have also
been
computed by Yin \& Miller (1995), who stress the dramatic effects that can
arise from photon trapping. But their solutions are only for stationary
flows.

\begin{figure}
\vspace{8.2cm}
\includegraphics{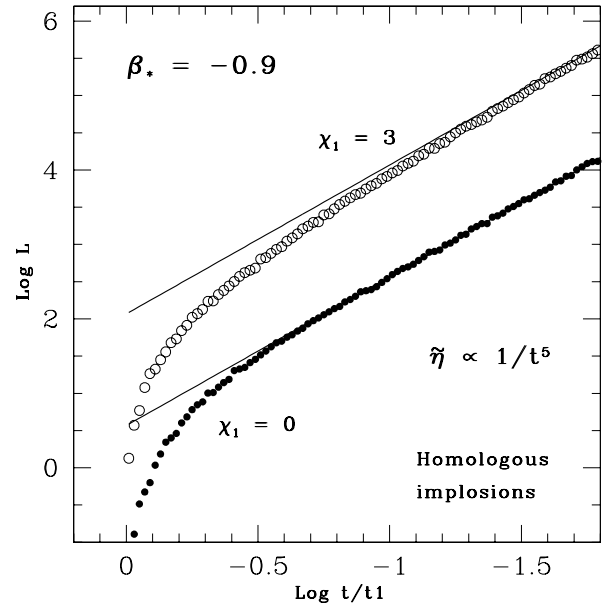}
\caption{Implosions. Same as Fig. 1 but with signs reversed for
$\beta_{*}$ and $t_{1}$. The luminosites for $\chi_{1} = 3$ have been
increased by $1.5dex.$ }
\end{figure}

\section{Testing a transfer code}

Apart from the degenerate case $\chi_{1}=0$, the exact solution of Sect. 3
is not appropriate for the precision testing of a conventional relativistic
transfer code since the implied scale-free emissivity $\eta_{1}$ could
only be determined approximately. Accordingly,
we now seek exact solutions when $\eta_{1}$ is specified rather than
$\tilde{\eta}_{1}$.

\subsection{Solution along characteristics}

	As did Mihalas (1980, Sect. IIIb) for his general equation,
we construct characteristics for Eq.(7) such that the partial differential
operator becomes a perfect differential. Thus Eq. (7) becomes
\begin{equation}
 \frac{1}{ct_{1}} \frac{d I_{1}}{ds} +
 \left[ \chi_{1} -\frac{(p - 4) \, \gamma +
                   p \, \gamma \beta \, \mu}{ct_{1}} \right]
  \, I_{1} =  \eta_{1} 
\end{equation}
where $c t_{1} s$ is distance along a characteristc defined by the
equations
\begin{equation}
  \frac{d \beta}{ds} = \frac{\mu}{\gamma} \;\;\;\;\;\;\; and \;\;\;\;\;\;\;
 \frac{d \mu}{ds} = \frac{\gamma}{\beta} (1 - \mu^{2})
\end{equation}
Dividing these equations, we find that $\gamma \beta \sqrt (1- \mu^{2})$
is constant along characteristics. Accordingly, the family of characteristics
is given by 
\begin{equation}
  \gamma \beta \sqrt (1- \mu{^2}) =
            \gamma_{*} \beta_{*} \sqrt (1- \mu_{*}^{2}) = \alpha
\end{equation}
where $ -1 < \mu_{*} < 0$ is a characteristic's direction cosine
at its entry point into the configuration.
Integrations of Eq. (15) proceed inwardly
along such characteritics starting at $s=0$ with the boundary condition
$I_{1}(\mu_{*};\beta_{*}) = 0$ and continuing until the characteristic
emerges at the surface with $\mu = |\mu_{*}|$.

\begin{figure}
\vspace{8.2cm}
\includegraphics{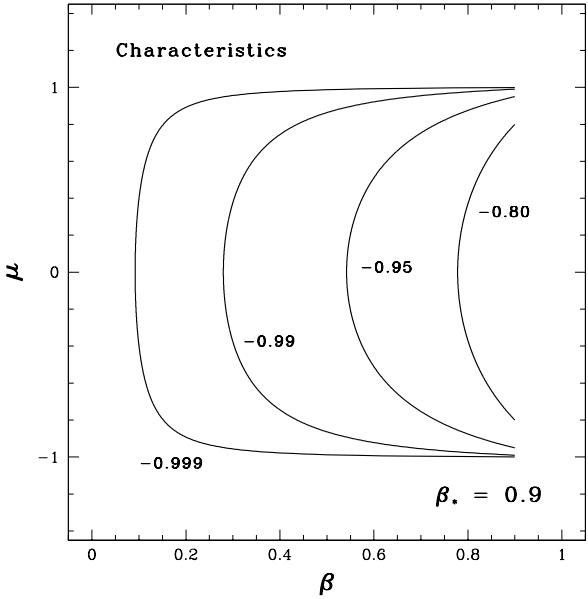}
\caption{Characteristic trajectories $\mu(\beta; \mu_{*})$
for Eq. (7) when $\beta_{*} = 0.9$. The curves are
labelled with the value of $\mu_{*}$, the direction cosine at the surface
for the inward ray.}
\end{figure}

	In Fig. 3, the characteristics are plotted for various values
of $\mu_{*}$ when $\beta_{*} = 0.9$. Note that, to obtain a
characteristic that penetrates close to the centre,
the parameter $\mu_{*}$ must closely approach $-1$. The point of closest
approach
($\mu = 0$) is at 
$\gamma =  \sqrt{1+ \alpha^{2}} = \gamma_{0} \,$,
or, equivalently, at $\beta = \alpha/\gamma_{0} = \beta_{0}$.

	Parenthetically, we note that these analytic characteristic curves
provide a
further powerful test of the relativistic
MC code described in Sect. 3.3. In the free streaming case,
each ${\cal E}$-packet follows its appropriate characteristic to high
precision, as it should. At creation, an ${\cal E}$-packet's initial $\beta$
and $\mu$
determine its invariant $\gamma \beta \sqrt (1- \mu{^2})$
and hence $\mu_{*}$ from Eq. (17). The packet then propagates along this
characteristic in the direction of increasing $s$ until
it escapes at the surface $\beta = \beta_{*}$ with $\mu = |\mu_{*}|$.

	The simplicity of the characteristics for homologous flow
allows the dimensionless arc length $s$ to be evaluated analytically as a
function of $\beta$.
Integrating the first member of Eq. (16) after eliminating 
$\mu$ with Eq. (17), we find that, along the inwardly-directed $(\mu < 0)$ 
segment of a characteristic,
\begin{equation}
  s(\beta; \mu_{*}) = \frac{1}{\gamma_{0}} \, \left[  sin^{-1} 
 (\frac{\gamma_{0}}{\gamma}) - sin^{-1}
 (\frac{\gamma_{0}}{\gamma_{*}}) \, \right]
\end{equation}
The corresponding formula after the point of closest approach $(\mu > 0)$
is
\begin{equation}
  s(\beta; \mu_{*}) =  \frac{1}{\gamma_{0}} \,        \left [ \, \pi - 
sin^{-1} (\frac{\gamma_{0}}{\gamma}) -
                      sin^{-1} (\frac{\gamma_{0}}{\gamma_{*}}) \, \right]
\end{equation}

	Because the function $s(\beta; \mu_{*})$ is readily
inverted to give $\beta(s; \mu_{*})$, the remaining
quantities $\gamma, \mu, \eta_{1}(\beta)$ and $\chi_{1} (\beta)$ can
likewise be
transformed into functions of $s$ along the characteristic defined by
$\mu_{*}$ for the given $\beta_{*}$; and this remark therefore
applies also to the coefficient of $I_{1}$ in Eq. (15). 

	Let us now define the effective extinction coefficient along
a characteristic to be  
\begin{equation}
 \hat{\chi}_{1} (s; \mu_{*}) =
 \chi_{1} -\frac{(p - 4) \, \gamma +
                   p \, \gamma \beta \, \mu}{ct_{1}} 
\end{equation}
with corresponding effective optical depth 
\begin{equation}
 \hat{\tau}_{1} (s; \mu_{*}) = ct_{1}
                 \int_{0}^{s}  \hat{\chi}_{1} \, ds
\end{equation}

	In terms of these quantities, the formal solution of Eq. (15) 
subject to the boundary condition $I = 0$ at $s = 0$ is
\begin{equation}
 I_{1} (s; \mu_{*}) = ct_{1} \int_{0}^{s}
   \eta_{1}(s^{'}) \, 
                e^{\hat{\tau}(s^{'}) - \hat{\tau}(s)} \,d s^{'}
\end{equation}
	With straightforward numerical integrations, Eqs. (18) - (22) allow
the intensity $I(s; \mu_{*})$ to be determined as a function $s$ along the
characteristic defined by $\mu_{*}$. By varying $\mu_{*}$ from $-1$ to $0$, we
can thus determine $I$ throughout the $(\mu, \beta)$-plane.

	This reduction of the problem to solution with the formal integral 
strongly suggests that these results could also be obtained with the
method developed by Baschek et al. (1997). Although their paper is
restricted to stationary relativistic flows, they note that
time-dependent problems can be treated.

\subsection{Equivalent static medium}

	The above analysis shows that, with the scaling
assumptions for $\eta$ and $\chi$ given in Eqs. (5) and
(6), the time-dependent relativistic transfer problem for homologous spherical
flow, reduces asymptotically to a tranfer problem 
in a static medium. For explosions, the solution tends to
this asymptote
as $t \rightarrow \infty$; for implosions, the limit is 
$t \rightarrow 0$. For free-streaming, the similarity solution is achieved
after all light signals emitted from within the configuration at $t = t_{1}$
have escaped.

	Because light travel-time and relativistic effects
are absent in a static medium, these effects
reappear in Eq. (20) as corrections to the extinction coefficient.
The term $\propto (p-4)$, which survives in the limit
$\beta \rightarrow 0$, represents the combination of finite propagation 
speed and the time-dependence of the emissivity. The second
correction term, which $\rightarrow 0$ as $\beta \rightarrow 0$, 
represents relativistic effects.

	Note that these corrections can give $\hat{\chi}_{1} < 0$, which
implies that $\hat{\tau}_{1}$ is not necessarily a monotonically
increasing function of $s$. Accordingly, $\hat{\tau}_{1}$ is not
appropriate as an alternative independent variable for Eq. (15).

\subsection{A particular case: moment solution}

Although the analysis of Sect. 4.1 provides a complete solution for the
scale-free radiation field without assumptions about $\eta_{1}(\beta)$
or $\chi_{1}(\beta)$, the answer is not in closed analytic form as was
the earlier result (Eq. [13]) for the cmf flux $H_{1}$ when the net emissivity
$\tilde{\eta}$ is independent of $\beta$. Interestingly, an analogous
result can be constructed when the conventional emissivity $\eta$
is the quantity specified.

	The coefficient of $J_{1}$ in Eq. (8) is zero if we set
\begin{equation}
 \chi_{1}(\beta) = (p - 4) \, \frac{\gamma}{c t_{1}}
\end{equation}
Thus $p > 4$ is necessary for $\chi_{1} > 0$, while $p = 4$ gives
free-streaming radiation.

	With $\chi_{1}$ given by Eq. (23),  
Eq. (8) simplifies to
\begin{equation}
 \frac{d H_{1}}{d \beta} +
 \frac{\gamma^{2}}{\beta} (2- p \, \beta^{2}) \, H_{1} 
 = c t_{1} \, \gamma \, \eta_{1} 
\end{equation}
This equation has integrating factor $\beta^{2}/\gamma^{p-2}$, and so
the solution satisfying the boundary condition $H_{1}(0) = 0$ is
\begin{equation}
 H_{1}(\beta) = c t_{1} \, \frac{\gamma^{p-2}}{\beta^{2}}
      \int_{0}^{\beta}  \eta_{1}(b) \; b^{2} (1 - b^{2})^
                        {\frac{p-3}{2}} \; d b 
\end{equation}
	A simple case for which this integral is analytic is obtained by 
assuming that $\eta_{1}$ is independent of $\beta$ and that $p = 5$. 
The result is
\begin{equation}
 H_{1}(\beta) = \frac{1}{15} c t_{1} \, \eta_{1} \, \gamma^{3} \beta
   \, (5 - 3 \beta^{2}) 
\end{equation}
With this exact formula, a code that solves relativistic transfer problems
with conventional techniques can be subjected to tests similar to those
described in Sect. 3.3 for a MC code based on ${\cal E}$-packets.

\subsection{A particular case: complete solution}

The particular case of Sect. 4.2 is remarkable in that the complete solution
can be obtained in closed form. The starting point is Eq. (15) with
$\chi_{1}$
from Eq. (23). The independent variable $s$ is conveniently transformed 
to $\beta$ using the first member of Eq. (16). The resulting transfer
equation is
\begin{equation}
 \frac{d I_{1}}{d \beta} - p \gamma^{2} \beta \, I_{1}
                    = ct_{1}\eta_{1} \, \frac{\gamma}{\mu}
\end{equation}
where, from Eq. (17),
\begin{equation}
 \mu( \beta) = \pm \frac{\gamma_{0}}{\beta} \sqrt{ \beta^{2} - \beta^{2}_{0}} 
\end{equation}
Here the negative root applies along the inward segment of a characteristic
as $\beta$ decreases from $\beta_{*}$ to $\beta_{0}$ - see Fig. 3. 
Thereafter, the positive root applies.

	The integrating factor for Eq. (27) is $\gamma^{-p}$. Accordingly,
substituting for $\mu$, imposing boundary condition
$I^{-}(\beta_{*}) = 0$, setting $p = 5$, and assuming that
$\eta_{1}$ is independent of $\beta$, we find that
the intensity of inwardly-directed radiation is

\begin{equation}
 I^{-}(\beta ; \mu_{*}) = c t_{1} \eta_{1} \frac{\gamma^{5}}{\gamma_{0}}
                 \int_{\beta}^{\beta_{*}}
           \frac{(1 - b^{2})^{2}} {\sqrt{b^{2}-\beta_{0}^{2}}} \, b \,d b
\end{equation}
which simplifies to
\begin{equation}
 I^{-}(\beta ; \mu_{*}) = c t_{1} \eta_{1} \, \frac{\gamma^{5}}{\gamma_{0}}
                      [ \, g(\beta_{*}; \mu_{*}) - g(\beta; \mu_{*}) \,]
\end{equation}
where
\begin{equation}
   g(\beta; \mu_{*}) = \xi \,  \left( \frac{1}{\gamma^{4}} + \frac{4}{3}
             \frac{\xi^{2}}{\gamma^{2}} +\frac{8}{15} \xi^{4}
                                              \right) 
\end{equation}
with
\begin{equation}
 \xi(\beta; \mu_{*}) =  \sqrt{\beta^{2} - \beta^{2}_{0}}
\end{equation}

	The outwardly-directed intensity is obtained
similarly. In this case, the boundary condition is
$I^{+}(\beta_{0}; \mu_{*} ) = I^{-}(\beta_{0}; \mu_{*} )$, and the 
solution is
\begin{equation}
 I^{+}(\beta ; \mu_{*}) = c t_{1} \eta_{1} \, \frac{\gamma^{5}}{\gamma_{0}}
                      [ \, g(\beta_{*}; \mu_{*}) + g(\beta; \mu_{*}) \,]
\end{equation}

	Eqs. (30) and (33) determine the intensity along
the characteristic defined by $\mu_{*}$. But these 
formulae can readily be used to calclulate $I_{1}(\mu; \beta)$, the intensity
as a function of $\mu$ at fixed $\beta$. Given $\mu$ and $\beta$,
Eq. (17) determines $\alpha$, which in turn determines
$\beta_{0}$, so that
$\beta_{0}= \beta_{0}(\mu, \beta)$ and similarly of course for $\gamma_{0}$.
Thus the right-hand sides of Eqs. (30) and (33) are now functions of
$\mu$ and $\beta$, and the characteristics are no longer relevant.

	This exact closed form analytic solution  
has been checked by substitution back into Eq. (7)
using numerical differentation to evaluate the two partial
derivatives.

	In Fig.3, the emergent intensity given by Eq. (33) is plotted
for various values of $\beta_{*}$. This shows extremely strong forward
peaking when $\beta_{*} = 0.9$, with huge intensities when $\mu \ga 0.9$. 
This is basically a light travel-time effect: with $\beta_{*} = 0.9$
radiation emerging with $\mu \sim 1$ includes photons emitted shortly after
the explosion when, since $j \propto 1/t^{6}$, the emissivity was much higher
than that now in the surface layers.

	One consequence of this forward
peaking is poor accuracy for Eddington's approximations. Thus, at the
surface of the $\beta_{*} = 0.9$ solution, $K_{1}/J_{1} = 0.821$ and
$H_{1}/J_{1} = 0.895$, as against Eddington's values of $1/3$ and $1/2$,
repectively.

\begin{figure}
\vspace{8.2cm}
\includegraphics{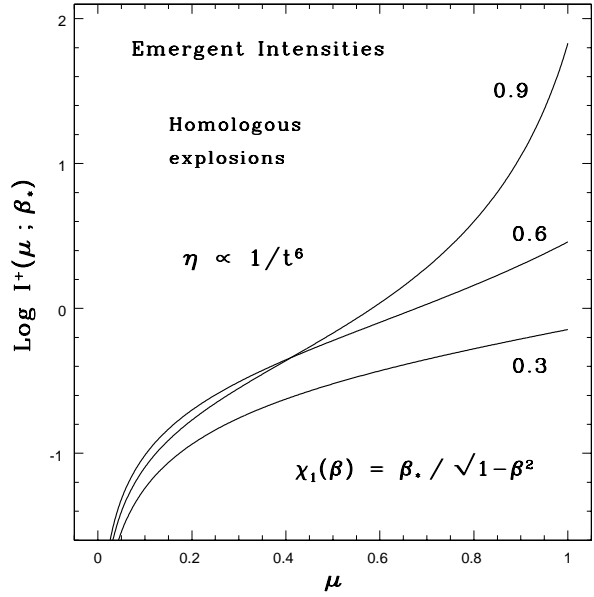}
\caption{Emergent intensities in the co-moving frame for the indicated
values of $\beta_{*}$.
The exponent $p = 5$, and $\chi_{1}$ is from Eq. (24). The formula for 
$\chi_{1}(\beta)$   
in units of $1/R_{1}$ is exhibited (cf. Figs.1 and 2). The unit of intensity
is $c t_{1} \eta_{1}$.}
\end{figure}

\subsection{A particular case: thermal emission}

Thus far the physical mechanisms responsible for extinction and emission
have not been specified. But let us now suppose that
\begin{equation}
 \chi = k + \sigma
\end{equation}
where $k$ and $\sigma$ are grey absorption and scattering coefficients,
respectively. 
The corresponding integrated emissivity coefficient is then
\begin{equation}
 \eta = kB + \sigma J
\end{equation}
where $B$ is the integrated Planck function, and the scattering is 
isotropic in the cmf.  

	Eqs. (34) and (35) are consistent with the scaling of Eqs. (5)
and (6) if $k$ and $\sigma$ are $\propto 1/t$ and if $B \propto 1/t^{p}$.
Given these scalings, the exact solution of Sect. 4.2 can now be applied
as follows: we set $p=5$ and choose the scale-free functions $k_{1}(\beta)$
and $\sigma_{1}(\beta)$ such that their sum satisfies Eq. (23). Then,
since $J_{1}(\beta)$ can be computed from Eqs. (30) and (33), the
implied scale-free function $B_{1}(\beta)$ can be derived from Eq. (35). 
Because the thermal emissivity is now known, this analytic solution can
be used to test a code that incorporates
an iteration procedure to detemine the scattering contribution to the
emissivity.  	

	The net emissivity 
$\tilde{\eta}_{1} = \eta_{1} - \chi_{1} J_{1} = k(B_{1} - J_{1})$ is
also determined by the above steps. Accordingly, this analytic solution
can be used to test MC codes based on ${\cal E}$-packets. Moreover,
this test is in principle more powerful than that provided by the flux
moment solution
of Sect. 3.2 since the angular distribution of the MC radiation field
can now also be checked.

\section{Conclusion}

The aim of this paper has been to derive exact analytic solutions in order
to test radiative transfer codes for relativistic flows. Such solutions
exist for spherically-symmetric homologous flows with power-law time
dependencies for the
grey extinction coefficient and the integrated emissivity. The
exact solution for the integrated intensity derived in Sect. 4.4, being
a function of three independent variables $t,\mu, \beta$ and one parameter
$\beta_{*}$, provides an extraordinarily demanding and informative test
for such codes. Moreover, its somewhat contrived derivation based
on a particular spatial variation of $\chi$ in no way 
lessens its usefulness. 

	In addition to deriving closed form analytic formulae for the
intensity and flux in the cmf, an exact formula has been derived for
the characteristics. This allows the kinematic and geometric aspects of
relativistic transport codes to be tested independently of the treatments
of absorption and emission. For MC codes, this
test is carried out by setting the extinction coefficient
to zero and then checking that photon packets propagate
along the known characteristics.

	In solving time-dependent transport problems with the RTE, the
time derivatives are commonly approximated with
a backward difference formula, thus using the solution at the previous
time step. This introduces errors $O(\Delta t)$ that
might well accumulate as the integration proceeds. With the availability
of an exact time-dependent solution, the magnitude of
the accumulated error can be determined. If the error accumulation is
unacceptable, a higher order difference formula can be employed 
that uses the solutions at the two previous time steps (Lucy 2004).

\end{document}